%% file: main.tex
\documentclass[sigconf]{acmart}

\renewcommand\footnotetextcopyrightpermission[1]{} % 去掉首页左下页脚的信
\usepackage{adjustbox}
\usepackage{multirow}

\AtBeginDocument{%
  \providecommand\BibTeX{{%
    \normalfont B\kern-0.5em{\scshape i\kern-0.25em b}\kern-0.8em\TeX}}}

\setcopyright{acmcopyright}
\copyrightyear{2024}
\acmYear{2024}
\acmDOI{XXXXXXX.XXXXXXX}

\acmConference[KDD '24]{ }{2024}{Conference}
\acmBooktitle{KDD '24}
\acmPrice{15.00}
\acmISBN{KDD '24}

\acmSubmissionID{498}

\begin{document}

\title{Cross-Domain LifeLong Sequential Modeling for Online Click-Through Rate Prediction}


\author{Ruijie Hou}
\affiliation{%
  \institution{Wechat Channels, Tencent}
  \city{Beijing}
  \country{China}
}
\email{jerriehou@tencent.com}

\author{Zhaoyang Yang}
\authornote{corresponding author}
\affiliation{%
  \institution{Wechat Channels, Tencent}
  \city{Guangzhou}
  \country{China}
}

\email{terrellyang@tencent.com}
\author{Yu Ming}
\affiliation{%
  \institution{Wechat Channels, Tencent}
  \city{Guangzhou}
  \country{China}
}
\email{julyeyang@tencent.com}

\author{Hongyu Lu}
\affiliation{%
  \institution{Wechat Channels, Tencent}
  \city{Guangzhou}
  \country{China}
}
\email{dreamlu@tencent.com}

\author{Zhuobin Zheng}
\affiliation{%
  \institution{Wechat Channels, Tencent}
  \city{Guangzhou}
  \country{China}
}
\email{jackzbzheng@tencent.com}

\author{Yu Chen}
\affiliation{%
  \institution{Wechat Channels, Tencent}
  \city{Guangzhou}
  \country{China}
}
\email{nealcui@tencent.com}

\author{Qinsong Zeng}
\affiliation{%
  \institution{Wechat Channels, Tencent}
  \city{Guangzhou}
  \country{China}
}
\email{qinzzeng@tencent.com}

\author{Ming Chen}
\affiliation{%
  \institution{Wechat Channels, Tencent}
  \city{Guangzhou}
  \country{China}
}
\email{mingchen@tencent.com}

\renewcommand{\shortauthors}{Ruijie and Zhaoyang, et al.}


\input{source/abstract}






\maketitle

\input{source/introduction}

\input{source/relatedwork}

\input{source/preliminary}

\input{source/methodology}

\input{source/experiment}

\input{source/conclusion}

\bibliographystyle{ACM-Reference-Format}
\bibliography{reference}

% %%
% %% If your work has an appendix, this is the place to put it.
% \appendix

\end{document}

%% file: source/abstract.tex
\begin{abstract}

Lifelong sequential modeling (LSM) has significantly advanced recommendation systems on social media platforms. Diverging from single-domain LSM, cross-domain LSM involves modeling lifelong behavior sequences from a source domain to a different target domain. In this paper, we propose the Lifelong Cross Network (LCN), a novel approach for cross-domain LSM. LCN features a Cross Representation Production (CRP) module that utilizes contrastive loss to improve the learning of item embeddings, effectively bridging items across domains. This is important for enhancing the retrieval of relevant items in cross-domain lifelong sequences. Furthermore, we propose the Lifelong Attention Pyramid (LAP) module, which contains three cascading attention levels. By adding an intermediate level and integrating the results from all three levels, the LAP module can capture a broad spectrum of user interests and ensure gradient propagation throughout the sequence. The proposed LAP can also achieve remarkable consistency across attention levels, making it possible to further narrow the candidate item pool of the top level. This allows for the use of advanced attention techniques to effectively mitigate the impact of the noise in cross-domain sequences and improve the non-linearity of the representation, all while maintaining computational efficiency. Extensive experiments conducted on both a public dataset and an industrial dataset from the WeChat Channels platform reveal that the LCN outperforms current methods in terms of prediction accuracy and online performance metrics.

\end{abstract}

%% file: source/introduction.tex
\section{Introduction}

Click-through rate (CTR) prediction stands as a fundamental task in many real-world applications.
% , which aiming to estimate the probability of users clicking on a presented item. 
The precision of CTR prediction heavily relies on comprehending the users' intentions towards the potential candidates. In recent years, deep neural networks (DNNs) have made significant strides in improving the accuracy of CTR prediction. They accomplish this by extracting representations of user interests from behavior sequences, specifically in relation to the candidate items \cite{cheng2016wide, zhang2019fieldaware, Wang_2021, din, dien}. 
% These advancements have played a crucial role in the rapid expansion of social media platforms.

Nowadays, social media platforms such as TikTok, YouTube, and WeChat Channels, presents billions of items to users every day. 
The complexity of user interactions on these platforms has significantly increased, posing new challenges in modeling user behavior sequences. 
On one hand, the volume of data has expanded exponentially, with some sequences extending to a length of lifelong. 
On the other hand, users often engage with a variety of content types, leading to behavior sequences that contains items from different domains.
This complexity becomes particularly pronounced when focusing on CTR prediction for smaller-scale domains.

\begin{figure}[t]
  \centering
  \includegraphics[width=0.9\linewidth]{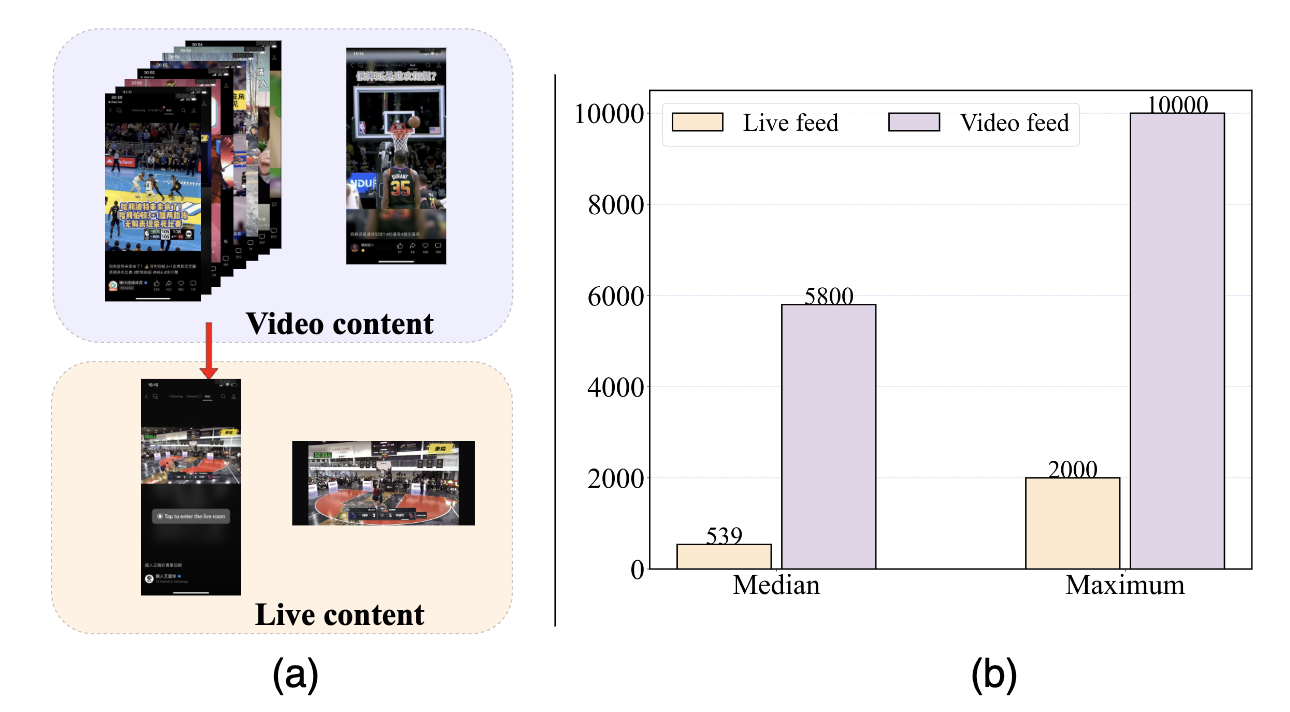}
  \caption{(a) A showcase of video content and live content in Wechat Channels platform. (b) A comparison between the statistics of the length of behavior sequence of video content and live content of users in Wechat Channels platform.}
  \vspace{-0.8cm}
  \label{fig:channels}
  \Description{}
\end{figure}

In order to overcome the lack of direct user data in the target domain, it becomes imperative for models to extract user interests from behavior sequences in an auxiliary source domain. 
For instance, as illustrated in \ref{fig:channels}, within WeChat Channels, the median behavior sequence length for live content is a mere 500, which is a fraction of the length for video content. 
Therefore, for live content CTR prediction, it is imperative to employ cross-domain lifelong sequential modeling (LSM) of video item interactions. This approach ensures comprehensive prediction that captures the user's integrated interests across the platform.

A common approach to managing lifelong sequences involves segmenting the modeling into two units: the General Search Unit (GSU) and the Exact Search Unit (ESU) \cite{qi2020searchbased}. The GSU's role is to sift through the sequence to identify items that are most relevant to the candidate items. Subsequently, the ESU is responsible for extracting user interest representations from the items identified by the GSU. This framework has been the foundation for numerous studies, leading to notable advancements in the field \cite{chen2021endtoend, cao2022sampling, chang2023twin}. 

Despite their successes, applying these methods to cross-domain LSM faces two primary challenges. Firstly, the effectiveness of GSU relies on expressive item embeddings to measure similarity between users' sequence items and candidate items. However, only optimizing these embeddings based on supervision signals from the target domain severely restricted their generalizability across domains and makes it difficult to capture cross-domain relations accurately. Aligning the representation space between candidate items in target domain and the sequence items in source domain becomes challenging. Secondly, most ESU methods use simpler attention techniques to minimize computational demands and maintain consistency with the GSU. However, cross-domain sequence contains more noise, which highly increases the difficulty and complexity of modeling. Therefore, more advanced attention techniques are necessary to effectively filter out irrelevant data within the sequence and enhance the final user interest representations in cross-domain LSM.

To address these challenges, we propose a novel Lifelong Cross Network (LCN) for cross-domain LSM. 
The LCN features a Cross Representation Production (CRP) module to improve the model's capabilities for item similarities across domains. Our experience suggests that users' short-term behaviors from different domains indicate similar intrinsic interests, which guides us in designing auxiliary supervision tasks. Inspired by contrastive learning, the CRP selects positive and negative item pairs from user behavior sequences within and across domains. This approach provides additional supervision on the item embeddings, which enhances the learning of relationships between items across domains.
Additionally, we integrate a Lifelong Attention Pyramid (LAP) module to enhance the extraction of user interest representations from lifelong sequences. The LAP module consists of three levels of cascading attentions, with each level processing top-ranked items from the previous level. The final output of LAP combines the results from all three levels, providing a comprehensive representation of user interests. This structure ensures full gradient propagation throughout the sequence, improving consistency across levels. This consistency further narrows the item pool of the top attention level, enabling the incorporation of advanced attention techniques to filter out noise in cross-domain lifelong sequences and enhance the non-linearity of representations while maintaining computational efficiency.

We conducted comprehensive experiments using both a public dataset and an industrial dataset collected from user interactions on the WeChat Channels platform. The results demonstrate that the proposed LCN outperforms existing methods in terms of prediction accuracy. The integration of CRP and LAP modules within the LCN has shown applicability to different LSM backbones and scenarios. Notably, the LCN has also achieved significant improvements in the A/B testing phase for online live content recommendations, with a relative increase of +2.93\% in CTR and +3.27\% in stay time. These results highlight the effectiveness and robustness of the proposed LCN in enhancing CTR prediction in cross-domain LSM.

\vspace{-2em}

%% file: source/relatedwork.tex
\section{RELATED WORK}
The modeling of user behavior sequences plays a central role in understanding user intentions, and extensive research has been dedicated to this field.
It has been observed that models tend to achieve better performance when longer sequence lengths are incorporated \cite{MIMN}.
With the exponential growth of data, the concept of lifelong sequential modeling (LSM) has emerged, aiming to extract user interests with respect to the candidate items over extensive sequence lengths. One notable approach addressing LSM is SIM \cite{qi2020searchbased}, which divides the process into a General Search Unit (GSU) and an Exact Search Unit (ESU), thereby managing lifelong sequences with reduced computational demands. Subsequent research has built upon this framework, leading to notable advancements \cite{Qin_2020, chen2021endtoend, cao2022sampling, chang2023twin}. However, these methods have not fully addressed the unique challenges associated with cross-domain LSM, and may experience performance declines due to insufficient transfer from the source to the target domain.

There are a diverse array of studies contributing to the field of cross-domain recommendation (CDR). A significant body of research advocates for the training of independent models within a source domain \cite{Hu_2018, liu2020cross, Minet}. Additionally, the variational auto-encoder (VAE) framework has been employed in several work to learn domain-invariant embeddings \cite{10.1145/3460231.3474265, liu2022exploiting, cao2022crossdomain}. Another approach involves the integration of Meta Networks \cite{Zhu_2021, zhu2021personalized}. Sequential modeling has also seen the incorporation of CDR techniques. Notably, PSJNet \cite{sun2021parallel} and $\pi$-Net \cite{10.1145/3331184.3331200} have adopted gating mechanisms to achieve this integration. In parallel, DA-GCN \cite{chen2019dagcn} and C2DSR \cite{cao2022contrastive} have leveraged Graph Neural Networks (GNNs), and RecGURU \cite{Li_2022} has utilized Transformer architectures. However, most of these studies focus on short sequence modeling, which limit their applicability to LSM due to the inherent differences in sequence length.

Contrastive learning has seen significant advancements in the fields of computer vision and natural language processing \cite{devlin2019bert, chen2020exploring, pathak2016context, SimCLR, chen2020simple}. It has also been adapted to enhance the Click-Through Rate (CTR) prediction tasks \cite{ma2020disentangled, sun2019bert4rec, xin2020selfsupervised, yao2021selfsupervised, zhou2021contrastive, liu2023at4ctr, wang2023cl4ctr, cao2022contrastive}. In the context of sequential modeling, MISS \cite{guo2022miss} utilizes contrastive learning to refine interest representation, while AQCL \cite{pan2021clickthrough} introduces an auxiliary loss to learn item relationships from sparse training samples. CL4CTR \cite{wang2023cl4ctr} aims to enhance item representation quality through contrastive learning, yet it requires the management of a large embedding table and triplets during training, which may be impractical in real-world applications. In this paper, we harness the principles of contrastive learning to impose additional supervision, enabling the model to learn item embeddings that can facilitate the bridging of items across different domains.

%% file: source/preliminary.tex
\section{PRELIMINARIES}

Cross-domain LSM involves modeling sequences of user behaviors from one domain (the source) and applying results gained to a different domain (the target). The objective of cross-domain LSM is to extract a representation of user interests based on their lifelong sequences in the source domain, which can then be used to enhance the accuracy of click-through rate (CTR) predictions in the target domain. A key aspect of this scenario is that the same user base is active in both domains, yet there is no item overlap between them.

Specifically, to formalize the modeling, we consider three distinct categories of features for each user:

\begin{itemize}
 \item Basic profile features, denoted as ${\{\beta\}}$. 
 \item Short-term behavior sequences in the target domain and the source domain, represented by $\vec{HT}^{t} = \{ ht_1, ht_2, \cdots, ht_{N} \}$ and $\vec{HS}^{t} = \{ hs_1, hs_2, \cdots, hs_{N} \}$ respectively. 
 \item The lifelong behavior sequence in the source domain, represented by $\vec{LHS}^{t} = \{ lhs_1, lhs_2, \cdots, lhs_{M} \}$. 
\end{itemize}

For a given user-item pair <$u_{i}$, $v_{i}$>, where $v_{i}$ is an item from the target domain, the model aims to predict the click-through rate (CTR) for user $u_{i}$ with respect to item $v_{i}$ as follows:
\begin{equation} 
    p_{i} = P(y_i=1|~u_{i},v_{i},\beta,\vec{HT}^{t},\vec{HS}^{t},\vec{LHS}^{t};\theta),
\end{equation}
where $\theta$ represents the parameters of the model.

The main network is optimized using a cross-entropy loss function, which is defined as follows:
\begin{equation}
\begin{split}
    \mathcal{L}_{CTR} = - \frac{1}{B} \sum_{i=1}^{B} & ~ \Big(~ y_i*log(p_i) + (1-y_i)*log(1-p_i) \Big),
\end{split}
\label{eq:cross_entropy}
\end{equation}
where $y_{i} \in \{0, 1\}$ represents the actual user feedback, and $B$ denotes the total number of user-item sample pairs in a training batch.

%% file: source/methodology.tex
\section{METHODOLOGY}

In this paper, we propose a novel approach, the Lifelong Cross Network (LCN) for cross-domain LSM. The LCN is comprised of two major components: the Cross Representation Production (CRP) module and the Lifelong Attention Pyramid (LAP) module.

The CRP module is a jointly trained sub-network with the objective of learning item embeddings that can bridge items across domains. These embeddings are then utilized within the main network to enhance its ability to identify the most relevant items with respect to a given candidate item from a lifelong sequence in the source domain. The LAP module is structured with three levels of cascading attentions, each processing the top ranked items of the previous level. The progressive nature of the LAP module ensures a more context-aware extraction of interest representations that are directly relevant to the target item. 

\begin{figure*}[ht]
  \centering
  \includegraphics[width=0.8\linewidth]{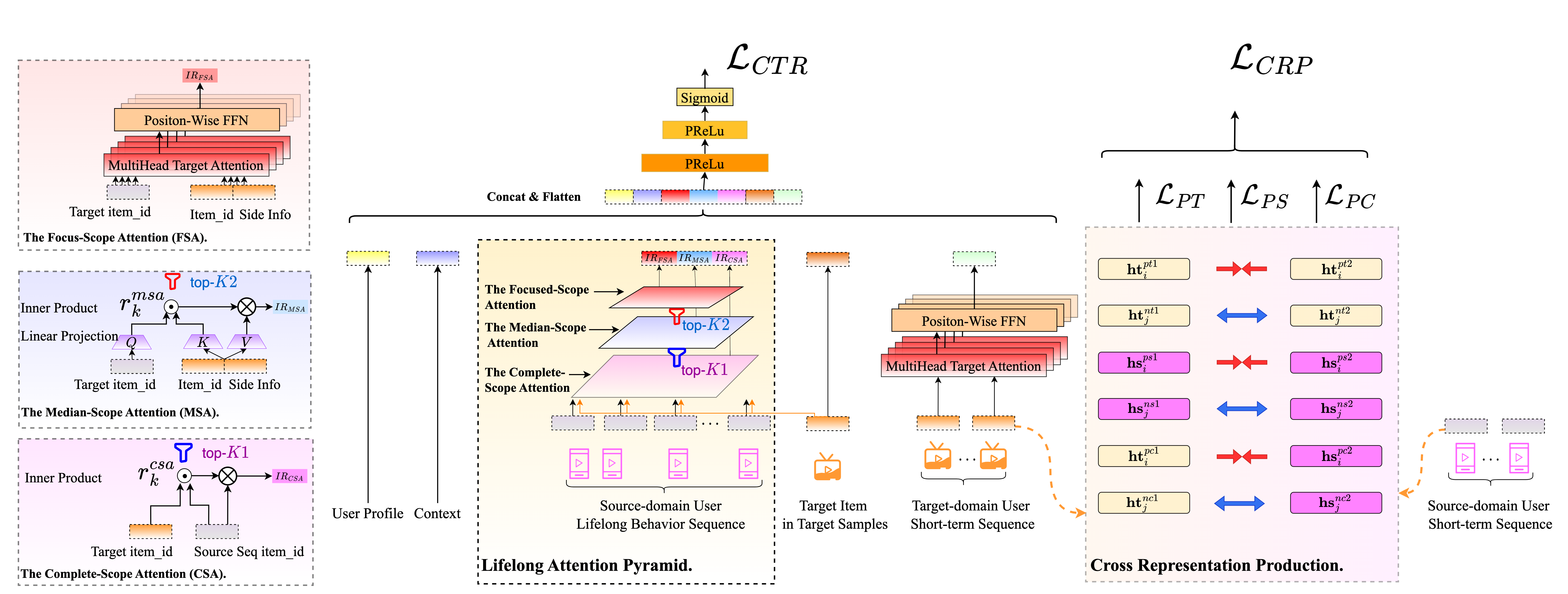}
  \caption{
  An overview of the proposed Lifelong Cross Network (LCN). There are two major components in the model: (i) The Cross Representation Production (CRP) module is a jointly trained sub-network to learn item embeddings capable of identifying similar items across domains. (ii) The Lifelong Attention Pyramid (LAP) module is composed of three levels of cascading attentions. These attentions are designed to progressively extract interest representations with respect to candidate items from the cross-domain lifelong sequence.
  }
  \label{fig:cross-domain}
  \Description{}
\end{figure*}

We show an overview of the proposed LCN in Figure \ref{fig:cross-domain}. 

\subsection{Cross Representation Production}
\label{sec:CRP}

A lifelong behavior sequence records a user's interactions over an extended period. When considering a particular item, only a fraction of this sequence may hold predictive value for the user's click-through rate (CTR) on that item. This is particularly true for cross-domain lifelong sequences, where the items originate from a different domain than the target item. It is crucial for the model to identify the most relevant items within the sequence to optimize model capacity and computational efficiency. 

Typically, Lifelong Sequence Modeling (LSM) is segmented into two units: a General Search Unit (GSU) and an Exact Search Unit (ESU). The role of GSU is to sift through the lifelong sequence and identify the items most relevant to the candidate item. Its effectiveness highly depends on the quality of the item embeddings utilized. Previous approaches re-use item embeddings learned during training of the model, which are proved to perform well within the training data's distribution. However, when the candidate item and the sequence items belong to different domains, the embeddings must be developed to bridge the gap between source and target domains. Achieving this is non-trivial, given that the model is primarily trained on the target domain data.

To overcome this challenge, we propose a Cross Representation Production (CRP) module that jointly refines cross-domain item embeddings along with the main network. Drawing inspiration from contrastive learning, the CRP module pairs positive and negative examples from user's short-term behavior sequences $\vec{HT}^{t}$ and $\vec{HS}^{t}$. It then imposes extra supervision on the item embeddings to enhance learning of relationships between items across domains.

\subsubsection{Positive and Negative Sampling.}

The construction of sample pairs is fundamental to contrastive learning. Within our CRP module, we achieve this by selecting items from user's short-term behavior sequences. This selection process is grounded in the understanding that a user's interests tend to remain stable across different domains, especially within a short-term period of time. Consequently, items within a user's short-term behavior sequence are likely to exhibit similar characteristics, regardless of their domain of origin. This consistency enables us to uniformly sample both positive and negative pairs across domains. 

\begin{figure}[]
  \centering
  \includegraphics[width=\linewidth]{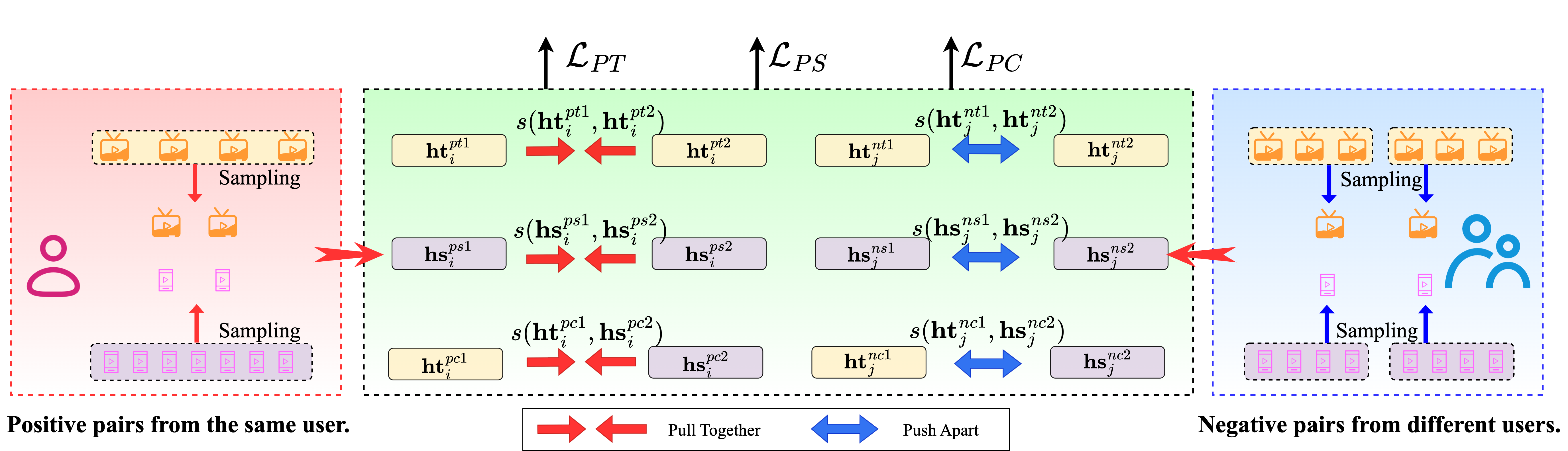}
  \caption{An illustration of the Cross Representation Production (CRP) module. It samples three distinct types of positive pairs and negative pairs. We introduce a contrastive loss to reduce the cosine distance between items in positive pairs while expand those of negative pairs.
  }
  \label{fig:CRP_overall}
  \Description{}
  \vspace{-0.3cm}
\end{figure}

As depicted in Figure \ref{fig:CRP_overall}, we have designed three distinct types of positive pairs for each user. To begin with, we select two items, $ht^{pt1}$ and $ht^{pt2}$, from the target domain's short-term sequence $\vec{HT}^{t}$, forming a positive pair within the target domain. Similarly, we choose two items, $hs^{ps1}$ and $hs^{ps2}$, from the source domain's short-term sequence $\vec{HS}^{t}$, creating a source domain positive pair. These pairs are intended to encourage the model to bring item embeddings of similar items closer within their respective domains. To learn alignment between similar items across domains, we construct a cross-domain positive pair by selecting one item each from $\vec{HT}^{t}$ and $\vec{HS}^{t}$, denoted as $ht^{pc1}$ and $hs^{pc2}$.

Regarding negative sample pairs, we employ a parallel approach to the positive sampling, but with items from different users' behavior sequences within the same training batch. This results in three types of negative sample pairs: <$ht^{nt1}$, $ht^{nt2}$>, <$hs^{ns1}$, $hs^{ns2}$> and <$ht^{nc1}$, $hs^{nc2}$>. For each type, we sample $M$ pairs per training batch, which are then used as the negative counterparts for all positive pairs within that batch.

\subsubsection{Loss Function}

The CRP module employs a contrastive loss function designed to enforce the model to minimize the cosine distance between item embeddings for positive pairs within a given batch. Each type of positive pairs is associated with a corresponding loss function as follows:

\begin{equation} 
\small
\label{ssl_loss_pt}
\mathcal{L}_{PT}= -\frac{1}{|B|} \sum_{i}^{B}\log \frac{\exp{(s(\mathbf{ht}_i^{pt1}, \mathbf{ht}_i^{pt2})/\tau)}}{\sum_{j}^{M} \exp{(s(\mathbf{ht}_j^{nt1},\mathbf{ht}_j^{nt2})/\tau)}}
\end{equation}

\begin{equation} 
\small
\label{ssl_loss_ps}
\mathcal{L}_{PS}= -\frac{1}{|B|} \sum_{i}^{B}\log \frac{\exp{(s(\mathbf{hs}_i^{ps1}, \mathbf{hs}_i^{ps2})/\tau)}}{\sum_{j}^{M} \exp{(s(\mathbf{hs}_j^{ns1},\mathbf{hs}_j^{ns2})/\tau)}}
\end{equation}

\begin{equation}
\small
\label{ssl_loss_pc}
\mathcal{L}_{PC}= -\frac{1}{|B|} \sum_{i}^{B}\log \frac{\exp{(s(\mathbf{ht}_i^{pc1}, \mathbf{hs}_i^{pc2})/\tau)}}{\sum_{j}^{M} \exp{(s(\mathbf{ht}_j^{nc1},\mathbf{hs}_j^{nc2})/\tau)}}
\end{equation}
where $B$ represents the size of training batch, and $s(x, y)$ denotes the cosine similarity between $x$ and $y$. 

The final loss of the CRP module, denoted as $\mathcal{L}_{CRP}$, is a integration of Equation \ref{ssl_loss_pt}, \ref{ssl_loss_ps} and \ref{ssl_loss_pc}:

\begin{equation}
\mathcal{L}_{CRP} = \lambda_{PT}\mathcal{L}_{PT} + \lambda_{PS}\mathcal{L}_{PS} + \lambda_{PC}\mathcal{L}_{PC}
\label{eq:lossS}
\end{equation}
where $\lambda_{PT}$, $\lambda_{PS}$ and $\lambda_{PC}$ represent the loss weights.

In the experimental section, we will demonstrate that integrating the CRP module substantially enhances the representational quality of the item embeddings. This improvement, in turn, boosts the final prediction accuracy across a broad spectrum of LSM frameworks.

\subsection{Lifelong Attention Pyramid}
\label{sec:LAP}

Classic LSM framework segments interest extraction into two units: the General Searching Unit (GSU) and the Exact Searching Unit (ESU), with the latter generally employing more complex attention techniques than the former. Although the GSU stage significantly narrows down the item pool for the ESU stage, the attention utilized in ESU are often less sophisticated than state-of-the-art (SOTA) attention techniques to maintain computational efficiency.

In the context of cross-domain LSM, items from the source domain in the behavior sequence and the candidate items from the target domain do not overlap, and the contextual and behavioral patterns can vary greatly between domains. Consequently, cross-domain lifelong sequences may introduce more noise, calling for better consistency across different searching stages and the use of more advanced attention techniques to interpret the sequences.

To address these challenges, we propose the Lifelong Attention Pyramid (LAP) module. This module extends the traditional two-stage framework into a three-level attention pyramid, featuring cascading levels of attention that aim to refine and streamline the search process within the lifelong sequence. By achieving better consistency across levels, the LAP reduces the number of items progressing to the top level. This reduction allows for the application of more advanced attention techniques at the top level to filter noise and enhance the non-linearity of the representation. We detail each of the levels, named the Complete-Scope Attention (CSA), the Median-Scope Attention (MSA), and the Focused-Scope Attention (FSA), in the following sections.

\subsubsection{The Complete-Scope Attention (CSA)} 

As the first level of attention, the Complete-Scope Attention (CSA) mirrors the function of the GSU from previous frameworks. Within the CSA, a broad yet general search is executed across the entire lifelong sequence. The objective is to ensure that every item within the sequence is accounted for, thereby excluding the least relevant items from subsequent levels. Meanwhile, the CSA can provide a preliminary interest representation by implementing a weighted average pooling based on the attention scores derived during the searching.

Given the considerable length of the input at this level and the less exact precision requirements, we employ a simple inner product calculation as the attention for the CSA. Formally, for a given candidate item $v_i$ and the lifelong behavior sequence $\vec{LHS}^{t} = \{ lhs_1, lhs_2, \cdots, lhs_{N} \}$, the attention score $r^{csa}_{k}$ for each item $lhs_k$ within the sequence is computed as follows: 
\begin{equation}
\small
r_{k} = e^h_{k} \odot e^v_{i}
\label{eq:score_CSA}
\end{equation}
where $e^h_{k}$ represents the embedding of the item $lhs_k$ in the sequence, and $e^v_i$ is the embedding of the candidate item. 

A straightforward weighted summation based on the score $r^{csa}_{k}$ is utilized to generate the interest representation output $IR_{CSA}$ of this level:
\begin{equation}
\small
IR_{CSA}= \sum_{k=1}^{N} r^{csa}_{k}*e^h_{k}
\label{eq:output_CSA}
\end{equation}

The top-$K_1$ items, as ranked by $r^{csa}_{k}$, are then selected to construct a reduced sub-sequence $\vec{LHS}^{c_1}$, which is subsequently fed into the second level of attention.

\subsubsection{The Median-Scope Attention (MSA)}

In contrast to previous two-stage methods, our approach introduces an intermediate Median-Scope Attention (MSA) level to effectively bridge the first and third levels. This addition is inspired by the observation that a significant majority of attention scores, often more than 90\%, tend to concentrate on a mere 20\% of the items in a sequence. By incorporating the MSA, we distribute some of the CSA's function of omitting less relevant items to this intermediate level, enhancing the overall consistency of the LAP module and further narrowing the item pool for the final attention level.

In the MSA, we incorporate additional contextual information about the items to perform a secondary attention search. It is important to note that the contextual details used here, such as interaction specifics like viewing duration, are consistent with the information that will be utilized in the final level. This consistency is crucial, as these contextual elements have been shown to significantly benefit sequential modeling \cite{chang2023twin}.

Formally, let $\vec{LCS}^{c_1}=\{ lcs_1, lcs_2,..., lcs_{K_1} \}$ represents the contextual information of the items provided by the CSA. The attention score $r^{msa}_{k}$ at this level is calculated using the following formulas: 
\begin{equation}
    r^{msa}_{k} = \frac{QK^\top}{\sqrt{d}},\ where\ Q=W^Qe^v_i,\ K=W^K(e^h_k || e^c_k),
\label{eq:score_MSA}
\end{equation}
where $W^Q$ and $W^K$ denote the attention weights, $d$ represents the inner dimension, $e^c_k$ is the embedding of the contextual information $lcs_k$ for item $lhs_k$, and $||$ signifies the concatenation operation.

The interest representation within the MSA, denoted as $IR_{MSA}$, is then derived through a weighted average pooling, expressed as: 
\begin{equation}
\small
IR_{MSA}= \sum_{k=1}^{K_1} r^{msa}_{k}*(W^V(e^h_{k}||e^c_{k}))
\label{eq:output_MSA}
\end{equation}
where $W^V$ is the projection matrix.

Following the first level, we form a sub-sequence $\vec{LHS}^{c_2}$ by selecting the top-$K_2$ items according to their $r^{msa}_{k}$ rankings. These selected items will then serve as the input for the final level.

\subsubsection{The Focused-Scope Attention (FSA)}

The objective of the final attention level mirrors that of the ESU, aiming to deliver a detailed and targeted interest representation with respect to the candidate item, based on the most relevant items filtered through the previous attention levels. The Focused-Scope Attention (FSA) benefits from a smaller set of items, enabling the use of a more advanced attention technique to increase the non-linearity of the representation.

We employ an attention similar to the decoder of the multi-head transformer \cite{vaswani2023attention} to extract interests from various perspectives. Formally, for a candidate item $v_i$ and the sub-sequence $\vec{LHS}^{c_2}$, the output of the $h$-th head in the multi-head attention is computed as: 
\begin{equation}
\begin{split}
    H_h &= Softmax(\frac{Q_hK_h^\top}{\sqrt{d}})V_h,\ where \\Q_h=W_h^Q&e^v_i,\ K_h=W_h^K(e^h_k || e^c_k),\ V_h=W_h^V(e^h_k || e^c_k)
\end{split}
\end{equation}
This attention process differs from the one used in MSA only in the multi-head aspect to maintain consistency. Building upon this, we integrate results of different head and use an additional feed-forward layer to further enhance the non-linearity of the model:
\begin{equation}
\begin{split}
    H=ReLU(concat(H_1,...,H_l)w_1+b_1)w_2+b_2
\end{split}
\end{equation}
where $l$ is the number of attention heads, $w_1$,$w_2$,$b_1$ and $b_2$ are the weights and biases of the feed-forward layer, respectively. The final interest representation, denoted as $IR_{FSA}$, is then obtained by:
\begin{equation}
\vspace{-0.2cm}
\begin{split}
    IR_{FSA}=\frac{1}{K_2}\sum_{k=1}^{K_2}(H_k)
\end{split}
\label{eq:output_FSA}
\end{equation}

At the end of the LAP module, we integrate the three interest representations, $IR_{CSA}$, $IR_{MSA}$ and $IR_{FSA}$, from each attention level to generate a representation that encapsulates a broad spectrum of user interests as reflected in the lifelong sequence. This integration also ensures full gradient propagation throughout the sequence, potentially enhancing the consistency between the different attention levels. The implications and benefits of this approach will be further explored and discussed in the experimental section.

The entire LCN, including the CRP module, is designed for end-to-end training. The final loss function for LCN is an combination of the CTR loss in Equation \ref{eq:cross_entropy} and the CRP loss in Equation \ref{eq:lossS}:

\begin{equation}
\vspace{-0.1cm}
\mathcal{L}= \mathcal{L}_{CTR} + \lambda_{CRP}\mathcal{L}_{CRP}
\label{eq:loss}
\end{equation}
where $\lambda_{CRP}$ denotes the factor to control the importance of ${L}_{CRP}$.

%% file: source/experiment.tex
\section{EXPERIMENTS}

To assess the performance of the proposed LCN, we carried out extensive experiments and compared it against previous methods using both a public dataset and an industrial dataset. Additionally, we implemented online A/B testing to further validate the effectiveness of LCN. The details of our experimental setup, along with the results and discussions, are introduced in this section.

\subsection{Experimental Settings}
\subsubsection{\textbf{Datasets}}

Our experiments was conducted using two datasets: one public dataset and one an industrial dataset. 

\textbf{Public Dataset}. We utilizes the Taobao Dataset\footnote{https://tianchi.aliyun.com/dataset/56}, which consists of traffic logs from Taobao's recommendation system. This dataset encompasses more than 2 million records from over 1 million users, collected over a period of 7 days. We have split the items in the the dataset into two distinct domains based on the ``Category ID'' associated with each item. Items with a ``Category ID'' below 2000 are allocated to the target domain, while those with a ``Category ID'' of 2000 or higher are allocated to the source domain. We consider the latest 24 user actions in both the source and target domains to construct short-term sequences. For lifelong sequences, we use the most recent 500 user actions from the source domain. The dataset is partitioned temporally, with the initial 6 days' data reserved for training and the data from the 7th day used as the test set.

\textbf{Industrial Dataset}. This dataset was collected from traffic logs of the Wechat Channels platform, containing user behavior sequences related to both video and live content. For each user, we collected his short-term behavior sequences for video and live items, as well as lifelong behavior sequence up to a maximum length of 2,000 for video items. The video item sequences were designated as the sequences in the source domain. The label of each sample is the user's click action with live items. The dataset comprises a collection of 2 billion records from 0.4 billion users, gathered over a period of 7 days. We partitioned the dataset temporally, with the data from the initial 6 days used for training and the data from the 7th day served as the test set.

\subsubsection{\textbf{Competitors}}

To evaluate the proposed LCN, we selected an array of state-of-the-art (SOTA) methods in the field of LSM for comparison. Our initial baseline was established using SIM Soft \cite{qi2020searchbased}. Below is an overview of the competing methods:

\begin{itemize}
  \item \textbf{SIM Soft} \cite{qi2020searchbased}: An early work that proposes to segment LSM into GSU and ESU stages. 
  \item \textbf{ETA} \cite{chen2021endtoend}: An approach for end-to-end lifelong sequential modeling that incorporates SimHash \cite{charikar2002similarity} for efficient GSU.
  \item \textbf{SDIM} \cite{cao2022sampling}: Introduces a novel Hash Attention technique to integrate GSU with ESU.
  \item \textbf{TWIN} \cite{chang2023twin}: An approach that employs dimension compression of cross features to enhance the consistency between GSU and ESU.
\end{itemize}

For training these models, we adhered to the parameter configurations recommended in their respective original publications or their open-source implementations.

\subsubsection{\textbf{Metrics}} For offline evaluation, we employed the following three key metrics: Area Under the Curve (AUC), Grouped Area Under the Curve (GAUC) and the value of the Logarithmic Loss (logloss) for the CTR prediction tasks. 

For online A/B testing, the evaluation was standardized by using the CTR and stay time of users on the items presented. We also monitored the inference latency as a secondary metric. 

\subsubsection{\textbf{Parameter Settings}}

The core architecture of our network consists of a straightforward two-layer feed-forward neural network. For the public dataset, the dimensions of the layers are configured to 256 and 128, while for the industrial dataset, we scale up to 512 and 256 to accommodate its larger size. The user profile features $\{B\}$ comprises profile information such as age, gender, and income level. We standardized the maximum length of the short-term behavior sequence to 50 for both datasets. All features, including item and user ids, are assigned distinct embedding spaces. A uniform embedding size of 64 is applied across all feature types. Model parameters are initialized using the Xavier Initialization method \cite{glorot2010understanding}, and the model is optimized with the Adam optimizer \cite{kingma2017adam} at a learning rate of 0.001. We set the batch size to 2048. Our models are developed using TensorFlow \cite{abadi2016tensorflow}. We trained the model using single A100 GPU for the public dataset, while 8 A100 GPUs are used a distributed manner for the industrial dataset.

\subsection{Module Analyses}

We detail experiments designed to assess the efficiency of the two major components of LCN. These experiments are carried out on the industrial dataset, which is more appropriate for our topic given its substantial data volume and extended sequence length.

\subsubsection{\textbf{Analyses of CRP}}

For the CRP module, we performed an ablation study alongside an embedding analysis to quantitatively and qualitatively assess its impact. Furthermore, we demonstrate the effectiveness of the CRP module when being incorporated with previous LSM frameworks.

\textbf{Ablation Study.} 
The outcomes of the experiments are presented in Table \ref{table:CRP_ablation}. In the experiment where the contrastive loss for the cross-domain sample pairs $\mathcal{L}_{PC}$ was omitted, the process of sampling cross-domain positive and negative pairs was also excluded. Additionally, in the case where no loss components were applied, the CRP module was entirely removed.

The results reveal that even without $\mathcal{L}_{PC}$, the CRP module contributes to a considerable enhancement over the initial baseline. This improvement suggests that the supervision provided by the contrastive loss aids the model in more effectively capturing the relationships between items within the embeddings, thereby refining the search quality in the GSU.

The best performance was achieved when all three contrastive losses were employed. This underscores the importance of incorporating an additional contrastive loss to learn the similarities between items across the source and target domains, which is crucial for cross-domain LSM. It appears that training the model solely with the target domain's CTR loss is insufficient for aligning embeddings across different domains. This misalignment is a primary factor in the degraded performance of previous LSM methods in cross-domain LSM, as the GSU struggles to deliver stable and precise search results without embeddings that encapsulate cross-domain information. The CRP module addresses this, significantly improve the efficiency of cross-domain LSM.

\begin{table}[t]
\footnotesize
\centering
\caption{Results of the ablation study of the CRP module.}
\vspace{-0.2cm}
\begin{tabular}{ccccc}
    \toprule
    \multicolumn{2}{c}{\textbf{Components of {$\mathcal{L}_{CRP}$}}} & \multicolumn{3}{c}{\textbf{Metrics}} \\
    \cmidrule(lr){1-2}\cmidrule(lr){3-5}
    $\mathcal{L}_{PT}$ \& $\mathcal{L}_{PS}$ & $\mathcal{L}_{PC}$ & AUC & GAUC & Logloss \\
    \midrule
    $\times$   & $\times$   &  0.7237 & 0.6379 & 0.2330 \\ % 
    \checkmark & $\times$   &  0.7250 & 0.6391 & 0.2290 \\ % 
    \checkmark & \checkmark &  \textbf{0.7294} & \textbf{0.6423} & \textbf{0.2223} \\ % 
    \bottomrule
\end{tabular}
\vspace{-0.4cm}
\label{table:CRP_ablation}
\end{table}

\begin{figure}[t]
  \centering
\hspace{-0.3cm}
    \includegraphics[width=0.38\textwidth]{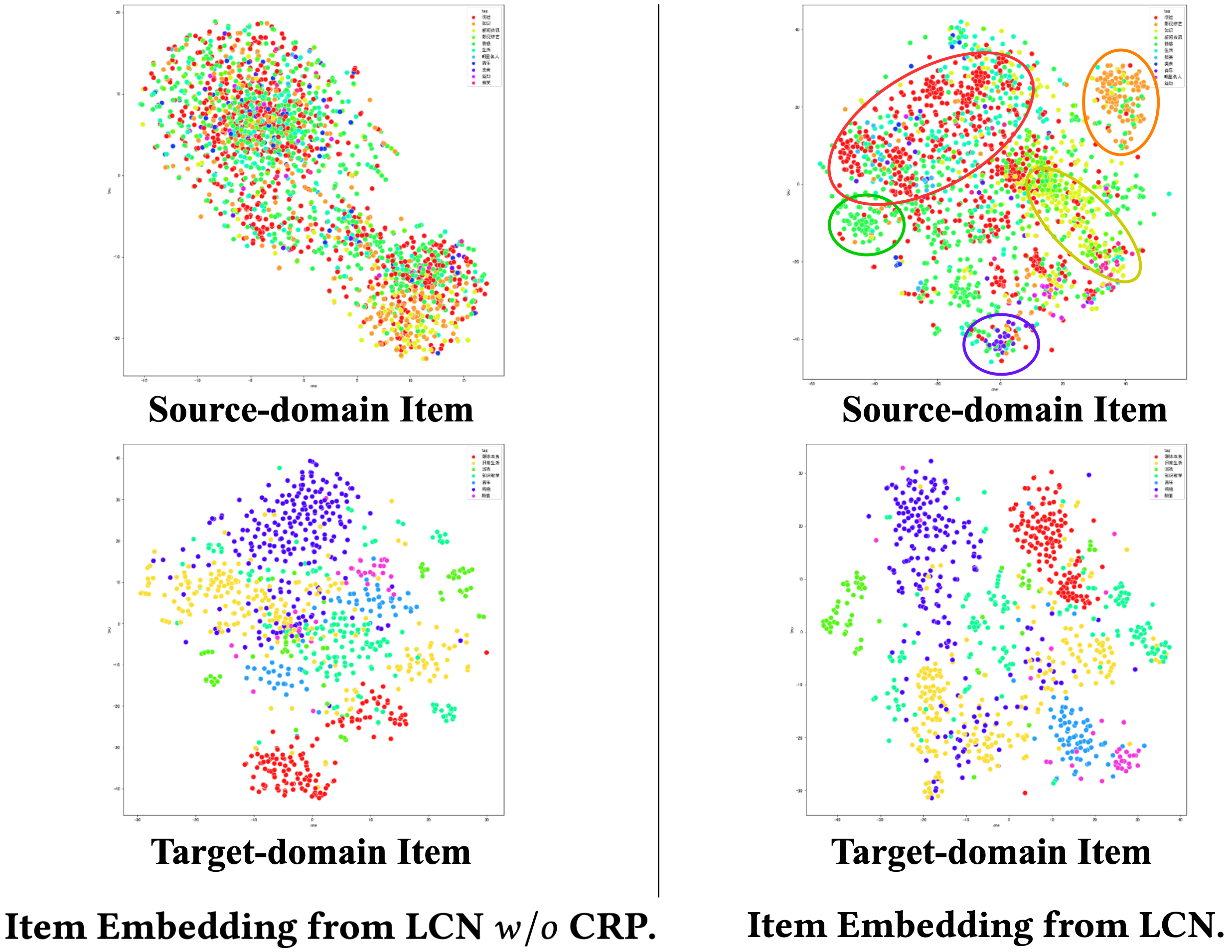}
  % \vspace{1px}
  \caption{Visualization of the item embeddings from $LCN\ w/o\ CRP$ and $LCN$. }
  \vspace{-0.5cm}
\label{fig:CRP_cluster}
\end{figure}

\textbf{Representation Quality.} 
To provide a clearer insight into the quality improvement of the learned item embeddings, we visualized the clustering of item embeddings, as depicted in Figure \ref{fig:CRP_cluster}. Specifically, we employed the T-SNE technique \cite{van2008visualizing} to distill the first two principal components from the learned embeddings, treating these values as coordinates in a two-dimensional space. By plotting these coordinates and color them according to item categories, we created a visual map that illustrates the embeddings' ability to differentiate between item categories, which is a direct measure of the embeddings' capacity to pair similar items.

We can observe that while embeddings learned without the CRP module can cluster items in the target domain with a quality comparable to those learned with the CRP module, there is a marked difference in the source domain, where embeddings refined with the CRP module exhibit superior clustering. This suggests that the additional contrastive loss introduced by the CRP module significantly enhances the quality of the embeddings in the source domain. This enhancement is a key factor in the performance improvements achieved by implementing the CRP module.

\textbf{Backbone Substitution.}
We extended our evaluation of the CRP module by integrating it into multiple LSM frameworks. In these experiments, the CRP module was utilized to refine item embeddings, which were then re-used in the GSU and ESU of the respective frameworks. The results are summarized in Table \ref{table:CRP_substitute}.

The results indicate that the CRP module consistently enhances the performance of all tested frameworks in managing cross-domain LSM. This improvement is largely attributed to the superior quality of the item embeddings generated for search purposes. These embeddings are proven to be beneficial regardless of variations in the underlying framework architecture.

It is important to note that in all experiments, the CRP module was trained jointly along with the main network. This means that while CRP introduces additional losses during the training phase, it does not impose any extra computational cost during inference.

\begin{table}[t]
    \centering
    \caption{Performance comparisons for different LSM methods when incorporating the CRP module.}
    \vspace{-0.3cm}
    \begin{adjustbox}{max width=0.6\linewidth}
    \begin{tabular}{cccc}
    \toprule
     Methods & AUC & GAUC & Logloss \\
    \midrule
    \multicolumn{1}{l|}{SIM Soft}         & 0.7196          & 0.6342            & 0.2356 \\
    \multicolumn{1}{l|}{SIM Soft$_{CRP}$} &\textbf{0.7255}  &\textbf{0.6375}    &\textbf{0.2252} \\
    \midrule
    \multicolumn{1}{l|}{ETA}              & 0.7166	        & 0.6328	        & 0.2373 \\
    \multicolumn{1}{l|}{ETA$_{CRP}$}      &\textbf{0.7204}	&\textbf{0.6365}	&\textbf{0.2214} \\
    \midrule
    \multicolumn{1}{l|}{SDIM}             & 0.7153	        & 0.6330	        & 0.2384 \\
    \multicolumn{1}{l|}{SDIM$_{CRP}$}     &\textbf{0.7192}	&\textbf{0.6358}	&\textbf{0.2191} \\
    \midrule
    \multicolumn{1}{l|}{TWIN}             & 0.7212	        & 0.6355	        & 0.2348 \\
    \multicolumn{1}{l|}{TWIN$_{CRP}$}     &\textbf{0.7265}	&\textbf{0.6391}	&\textbf{0.2259} \\
    \bottomrule
    \end{tabular}
    \end{adjustbox}  
    \vspace{-0.5cm}
\label{table:CRP_substitute}
\end{table}

\subsubsection{\textbf{Analyses of LAP}}
We carried out experiments to investigate various factors influencing the performance of the LAP, focusing on the consistency across different levels and the impact of level sizes. Additionally, we evaluated LAP in a single-domain LSM to demonstrate the applicability of the proposed LAP.

\textbf{Ablation Study.} 
A series of experiments was conducted to evaluate the influence of the LAP module, particularly examining its performance in relation to the size of the item pools for MSA ($K_1$) and FSA ($K_2$). The results are summarized in Table \ref{table:LAP_ablation}. In the absence of both $K_1$ and $K_2$, we established our initial baseline with the SIM Soft model. When either $K_1$ or $K_2$ was not included, the corresponding attention level was omitted from the LAP module.

The results indicate that LAP outperforms the initial baseline, even with the removal of either MSA or FSA. This improvement is largely attributed to the enhanced gradient flow across its attention levels, which improves the learning efficiency of the model. The performance further improves with the complete implementation of both MSA and FSA.

Furthermore, it can be observed that the performance exhibits minimal sensitivity to the specific values assigned to $K_1$ and $K_2$, particularly once they exceed a quarter of the size of the previous level. This confirms that the introduction of an intermediate MSA level promotes greater consistency across the levels. Consequently, the performance of the model remains stable, even with a reduced number of items progressing to the advanced attention in the FSA, which can contribute to decreased inference latency.

\begin{table}[t]
\footnotesize%\scriptsize
\centering
\caption{Comparison of model performance under different MSA and FSA settings.}
% \begin{adjustbox}{max width=1.0\linewidth}
\vspace{-0.2cm}
\begin{tabular}{cccccc}
    \toprule
    \multicolumn{2}{c}{\textbf{Size of the item pool}} & \multicolumn{3}{c}{\textbf{Metrics}} \\
    \cmidrule(lr){1-2}\cmidrule(lr){3-5}
    $K_1$ & $K_2$ &  AUC    & GAUC   & Logloss \\
    \midrule
    -     & -     &  0.7255	& 0.6375 & 0.2252  \\
    50    & -     &  0.7266	& 0.6388 & 0.2246  \\
    -     & 50    &  0.7297	& 0.6424 & 0.2222  \\
    \midrule
    100   & 30    &  0.7271 & 0.6396 & 0.2240  \\
    100   & 50    &  0.7279	& 0.6407 & 0.2234  \\
    \midrule
    200   & 30    &  0.7286 & 0.6415 & 0.2229  \\
    200   & 50    &  0.7294	& 0.6423 & 0.2223  \\
    200   & 100   &  0.7296 & 0.6424 & 0.2221  \\
    \midrule
    500   & 30    &  0.7288 & 0.6317 & 0.2228  \\
    500   & 50    &  0.7295	& 0.6424 & 0.2223  \\
    500   & 100   &  0.7297 & 0.6425 & 0.2220  \\
    \bottomrule
\end{tabular}
\vspace{-0.2cm}
% \end{adjustbox}
\label{table:LAP_ablation}
\end{table}

\textbf{Consistency Analyses.} 
We conducted a comparative analysis to examine the consistency across different stages of LSM among various LSM methods. For each method, we began by processing the entire lifelong sequence using the parameter and attention settings of the final LSM stage, ESU for previous methods and FSA for our proposed LAP. From this, we extracted the top-50 items based on their attention scores and used it as the real set. Similarly, we identified the top-50 items using the parameter and attention settings of the previous stages, GSU for previous methods and CSA, MSA for LAP, to obtain the test set. The consistency of each method was then evaluated by calculating the proportion of items in the real set that were identified in the test set. 

\begin{figure}[t]
\centering
\hspace{-0.3cm}
    \includegraphics[width=0.45\textwidth]
    {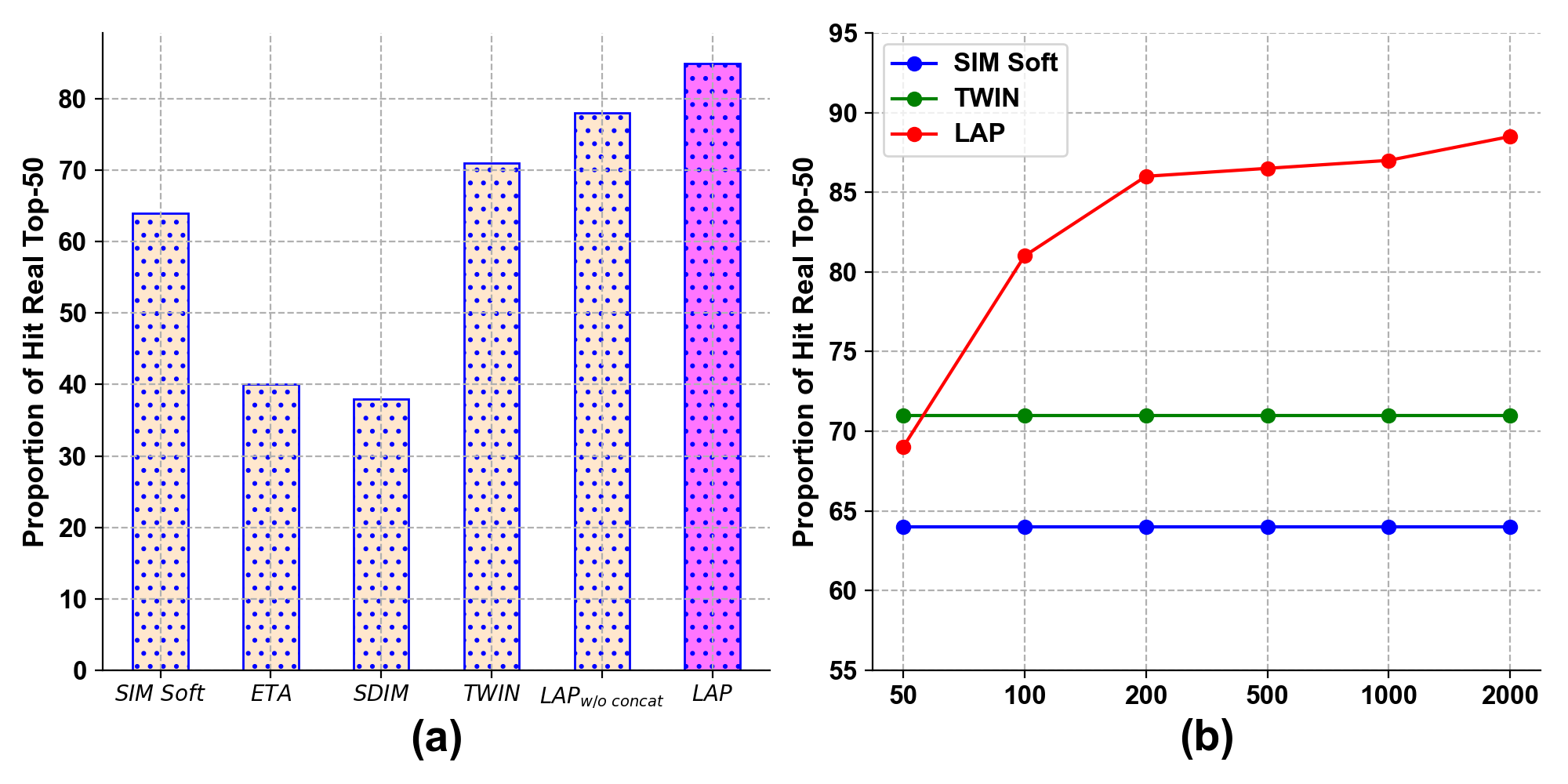}
  \vspace{-0.3cm}
  \caption{Comparisons of the consistencies between GSU \& ESU (a) of different LSM methods and (b) under different settings of $K_1$.}
\label{fig:LAP_consistency}
\vspace{-0.3cm}
\end{figure}

As depicted in Figure \ref{fig:LAP_consistency}-(a), the LAP method, with $K_1$ and $K_2$ set to 200 and 50 respectively, demonstrated superior consistency compared to other methods. This can be attributed to the presence of the MSA level, which effectively bridges the CSA and FSA. It is also worth noting that a significant enhancement was achieved when integrating (concatenating) the results from all three attention levels as the final output of LAP. This is mainly because that the integration facilitates gradient flow throughout the sequence, which is crucial for improving consistency across different levels.

We further conducted a set of experiments to assess the influence of 
$K_1$ on consistency. The results, depicted in Figure \ref{fig:LAP_consistency}-(b), reveal that consistency is augmented with larger values of $K_1$ and tends to plateau once $K_1$ exceed a quarter of the maximum sequence length, corroborating the model performance trends observed in the previous section. This indicates that the interplay between MSA and FSA is highly consistent, owing to the contextual information included in MSA. This also highlights the value of MSA, which is key to maintain consistency across levels while considerably narrowing the item pool needed in the FSA. 

\textbf{Adaptability Test.}
We conducted experiments to assess the adaptability of the proposed LAP in the context of single-domain LSM. For these experiments, we omitted CRP module and relied on user interactions within the original source domain (video content) as the training labels. The results are summarized in Table \ref{table:LAP_single}.

The results indicate that LAP continues to outperform previous methods in the context of single-domain LSM, though with a smaller margin of improvement compared to cross-domain LSM. This reduced gain is likely due to the relative ease of modeling lifelong sequences within a single domain, where the data distribution and behavioral patterns among items in the sequence and the candidate items tend to be more consistent.

\begin{table}[t]
\footnotesize%\scriptsize
\centering
\caption{Comparison of performance on single-domain LSM.}
% \begin{adjustbox}{max width=1.5\linewidth}
\begin{tabular}{cccccc}
    \toprule
    Methods       &     AUC &   GAUC & Logloss \\
    \midrule
    SIM Soft      &  0.7036	& 0.6495 & 0.2669  \\ 
    ETA           &  0.7030	& 0.6485 & 0.2692  \\ 
    SDIM          &  0.7025	& 0.6485 & 0.2688  \\
    TWIN          &  0.7044	& 0.6506 & 0.2664  \\
    LCN $w/o$ CRP &  \textbf{0.7076} & \textbf{0.6531} & \textbf{0.2648}  \\
    \bottomrule
\end{tabular}
% \end{adjustbox}
\label{table:LAP_single}
\vspace{-0.2cm}
\end{table}

\subsection{Overall Performance}

The overall performance results of the proposed LCN alongside previous methods on both public and industrial datasets are presented in Table \ref{table:LCN_overall}. We denote different settings of $K_1$ and $K_2$ as LCN-$K_1$-$K_2$. 

Across all metrics and datasets, the proposed LCN consistently outperforms the other methods. It is important to note, however, that the margin of improvement on the public dataset is narrower than that on the industrial dataset. This is largely due to the public dataset's smaller sequence lengths and data volume. The performance difference on the industrial dataset is likely a more accurate reflection of the model's true generalization capabilities and its effectiveness in real-world recommendation scenarios.

\begin{table}[t]
    \centering
    \caption{Final performance comparisons on both the public and the industrial datasets.}
    \begin{adjustbox}{max width=1.0\linewidth}
    \begin{tabular}{ccccccc}
    \toprule
        \multirow{2}*{Methods} & \multicolumn{3}{c}{Industrial.} & \multicolumn{3}{c}{Public.} \\
        \cmidrule(lr){2-4}\cmidrule(lr){5-7}
        & AUC & GAUC & Logloss  & AUC & GAUC & Logloss \\
        \midrule
        SIM Soft    & 0.7196 & 0.6342 & 0.2356 & 0.6071 & 0.5762 & 0.1629 \\
        ETA         & 0.7166 & 0.6328 & 0.2373 & 0.6055	& 0.5748 & 0.1630 \\
        SDIM        & 0.7153 & 0.6330 & 0.2384 & 0.6038	& 0.5734 & 0.1631  \\
        TWIN        & 0.7212 & 0.6355 & 0.2348 & 0.6087	& 0.5776 & 0.1623 \\
        \cmidrule(lr){1-7}
        DSS[2]       & 0.7242 & 0.6386 & 0.2227 & 0.6107 & 0.5790 & 0.1617 \\
        BERT4Rec[20] & 0.7253 & 0.6392 & 0.2224 & 0.6122 & 0.5797 & 0.1615 \\
        SQN[22]      & 0.7241 & 0.6381 & 0.2228 & 0.6100 & 0.5787 & 0.1618 \\
        SAC[22]      & 0.7242 & 0.6381 & 0.2228 & 0.6104 & 0.5790 & 0.1618 \\
        SSL[31]      & 0.7246 & 0.6386 & 0.2227 & 0.6111 & 0.5792 & 0.1617 \\
        CLRec[35]    & 0.7279 & 0.6398 & 0.2225 & 0.6123 & 0.5799 & 0.1614 \\
        AT4CTR[37]   & 0.7262 & 0.6385 & 0.2225 & 0.6117 & 0.5793 & 0.1616 \\
        CL4CTR[38]   & 0.7265 & 0.6386 & 0.2225 & 0.6115 & 0.5791 & 0.1616 \\
        CL4SRec[40]  & 0.7276 & 0.6395 & 0.2224 & 0.6122 & 0.5798 & 0.1615 \\
        \cmidrule(lr){1-7}
        
        LCN-200-50 $w/o$ CRP & 0.7237 & 0.6379 & 0.2330 & 0.6098 & 0.5786 & 0.1619  \\
        LCN $w/o$ LAP        & 0.7266 & 0.6401 & 0.2296 & 0.6128 & 0.5808 & 0.1615  \\
        LCN-200-50           & \textbf{0.7294} & \textbf{0.6423} & \textbf{0.2223} & \textbf{0.6143} & \textbf{0.5820} & \textbf{0.1610}  \\
        LCN-500-100          & \textbf{0.7297} & \textbf{0.6425} & \textbf{0.2220} & \textbf{0.6145} & \textbf{0.5820} & \textbf{0.1610} \\
        \bottomrule
    \end{tabular}
    \end{adjustbox}
    \vspace{-0.3cm}
    \label{table:LCN_overall}
\end{table}

\subsection{Online Testing}

To further validate the efficiency of the proposed LCN, we conducted an online A/B test to assess the quality of its recommendations. We utilized LCN-200-50 to maintain a balance between computational efficiency and performance. Over a seven-day period, we collected user feedback to calculate online metrics. The results were significant, with group B achieving a relative increase of \textbf{+2.93\%} in CTR and \textbf{+3.27\%} in stay time. Moreover, the inference latency for group B was only 3 ms longer than that of group A, which is a negligible trade-off considering the substantial gains in user experience delivered.

%% file: source/conclusion.tex
\section{CONCLUSIONS}

In this paper, we propose the Lifelong Cross Network (LCN) for cross-domain lifelong sequential modeling (LSM). LCN is composed of two components: the Cross Representation Production (CRP) module and the Lifelong Attention Pyramid (LAP) module. The CRP module is a sub-network that is supervised by contrastive loss to learn item embeddings that can bridge similar items across domains. The LAP module is structured with three levels of cascading attentions to extract interest representations from the lifelong sequence with respect to the candidate item. By integrating these two modules, the proposed LCN can effectively identify relevant items across domains, achieving a highly consistent and computational efficient interest extraction process within the lifelong sequence in the source and improving the CTR prediction in the target domain. 

The results of experiments on public and industrial datasets reveal that the proposed LCN significantly improves the model's capacity to handle cross-domain LSM. The results also indicate the adaptability of the CRP and LAP modules for their effectiveness when integrated with different LSM backbones or applied to LSM within a single domain. As for future work, we aim to refine the integration of the CRP module with the main network and investigate enhancements to the LAP module's inter-level connections.